\documentclass[11pt]{article}
\usepackage{calc}

\usepackage{amsmath}
\usepackage{amssymb}
\usepackage{amsfonts}
\usepackage{epsfig}
\usepackage{cite}
\usepackage{calc}
\usepackage{color}

\usepackage{hyperref}
\usepackage{times}
\usepackage{subfigure}
\DeclareGraphicsExtensions{.eps}
\usepackage[all,2cell]{xy} \UseAllTwocells \SilentMatrices

\newcommand{\Q}{{\mathbb Q}}

\newcommand{\ba}{\begin{array}}
\newcommand{\ea}{\end{array}}

\newcommand{\no}{\nonumber}
\newcommand{\beq}{\begin{equation}}
\newcommand{\eeq}{\end{equation}}

\newcommand{\bea}{\begin{eqnarray}}
\newcommand{\eea}{\end{eqnarray}}
\newcommand{\bean}{\begin{eqnarray*}}
\newcommand{\eean}{\end{eqnarray*}}

\newcommand{\bit}{\begin{itemize}}
\newcommand{\eit}{\end{itemize}}
\newcommand{\ben}{\begin{enumerate}}
\newcommand{\een}{\end{enumerate}}

\title{Approximately-Universal Space-Time Codes for the
Parallel, Multi-Block and Cooperative-Dynamic-Decode-and-Forward
Channels }
\date{}
\author{Petros Elia and P. Vijay Kumar
\thanks{Petros Elia and P. Vijay Kumar are with the Department of EE-Systems,
University of Southern California, Los Angeles, CA 90089 ({\tt
\{elia,vijayk\}@usc.edu}). This work was carried out while P. Vijay
Kumar was on a leave of absence at the Indian Institute of Science
Bangalore. This research is supported in part by NSF-ITR Grant
CCR-0326628 and in part by the DRDO-IISc Program on Advanced
Research in Mathematical Engineering. }}

\begin{document}
\maketitle \thispagestyle{empty}

\begin{abstract}
Explicit codes are constructed that achieve the
diversity-multiplexing gain tradeoff of the cooperative-relay
channel under the dynamic decode-and-forward protocol for any
network size and for all numbers of transmit and receive antennas
at the relays.

A particularly simple code construction that makes use of the
Alamouti code as a basic building block is provided for the single
relay case.

Along the way, we prove that space-time codes previously constructed
in the literature for the block-fading and parallel channels are
approximately universal, i.e., they achieve the DMT for any fading
distribution. It is shown how approximate universality of these
codes leads to the first DMT-optimum code construction for the
general, MIMO-OFDM channel.

\end{abstract}

\section{Introduction} \label{sec:introduction}

Cooperative relay communication is a promising means of wireless
communication in which cooperation is used to create a virtual
transmit array between the source and the destination, thereby
providing the much-needed diversity to combat the fading channel.

Consider a communication system in which there are a total of $N+1$
nodes that cooperate in the communication between source node $S$
and destination node $D$. The remaining $(N-1)$ nodes thus act as
relays.  We follow the literature in making the assumptions listed
below concerning the channel.  Our description is in terms of the
equivalent complex-baseband, discrete-time channel.

\bit

\item All nodes have a single transmit and single receive antenna and are assumed to
transmit synchronously.

\item The number of channel uses $T$ over which communication
takes place is short enough to invoke the quasi-static assumption,
i.e., the channel fading coefficients are fixed for the duration
of the communication,

\item We assume half-duplex operation at each node, i.e., at any
given instant a node can either transmit or receive, but not do
both.

\item The noise vector at the receivers is assumed to be comprised of i.i.d.,
circularly symmetric complex gaussian $\mathbb{C}\mathcal {N}
(0,\sigma^2)$ random variables.

\eit

\section{The DDF Protocol} \label{sec:DDF_protocol}

Under the DDF protocol, the source transmits for a total time
duration of $BT$ channel uses.   This collection of $BT$ channel
uses is partitioned into $B$ blocks with each block composed of $T$
channel uses.  Communication is slotted in the sense that each relay
is constrained to commence transmission only at block boundaries. A
relay will begin transmitting after listening for a time duration
equal to $b$ blocks only if the channel ``seen'' by the relay is
good enough to enable it to decode the  signal from the source with
negligible error probability. We explain in more detail.

\subsection{Notation and Expressions for the Received Signal}
\label{subsec:notation}

Initially, we will assume that each relay node has a single transmit
antenna.  The extension to arbitrary number of antennas is
straightforward.  We will similarly make the initial assumption that
the destination node has a single receive antenna.

It will be convenient at times to regard the source as the first
relay, i.e., $S \ \equiv \ R_1$ and the destination as the $(N+1)$th
relay, i.e., $D \ \equiv \ R_{N+1}$. The notation below is with
respect to a fixed channel realization that lasts for the $B$-block
duration.

Let $\underline{x}_b(n)$, $1 \leq b \leq B$, $n=1,2,\cdots,N$ denote
the $T$-tuple transmitted by the $n$th node during the $b$th block.
Since all nodes do not transmit in all blocks, we will make the
assignment $\underline{x}_b(n)=\underline{\varphi}$, where we regard
$\underline{\varphi}$ as the ``empty'' vector to handle the case of
no transmission.   In particular, the vectors $\underline{x}_b(1), \
b=1,2,\cdots,B$ denote the $B$ successive transmissions by the
source.

Let us assume that up until the end of the $(b-1)$th block, we
know which relays began transmitting and when.   We will assume
that once a relay has begun transmitting, it will keep on
transmitting thereafter until the end of the $B$th block.   Let
${\cal I}_k$ denote the set of indices of the relays that transmit
during the $k$th block, $k=1,2,\cdots ,B$.   We will refer to
${\cal I}_k$ as the $k$th activation set.   Clearly \bean {\cal
I}_1 & = & \{1\}
\\ {\cal I}_k & \subseteq & {\cal I}_{k+1}, \  \ 1 \leq k \leq
(b-2). \eean   We next proceed to determine for the relays not in
${\cal I}_{b-1}$, whether or not the time is right for them to begin
transmission during the $b$th block.  In other words, we will
determine ${\cal I}_b$ given $\{{\cal I}_k\}_{k=1}^{b-1}$. Since
${\cal I}_1$ is known, this procedure will allow us to recursively
determine the activation sets ${\cal I}_k$ for all $1 \leq k \leq
B$.

We will begin by first identifying the signal received by such a
relay during the $(b-1)$th block.  Let $\zeta_b$, $ 1 \leq b \leq
B$, denote the size of ${\cal I}_b$ i.e.,
\[ \mid {\cal I}_b \mid \ = \ \zeta_b .\]
  Clearly,
  \[ 1= \zeta_1 \leq \zeta_2 \leq \cdots \leq \zeta_{b-1} \leq N.\]
Let the elements of ${\cal I}_k, \ 1 \leq k \leq (b-1)$, be given by
\[
{\cal I}_k \ = \ \{1=m_1, \ m_2, \ \cdots, \ m_{\zeta_k} \}  .
\]
 We use $h(m,n)$ to denote the fading coefficient between the $m$th  and $n$th nodes.   Let $n
\not \in {\cal I}_{b-1}$ and \bean \underline{h}_k(n) & = &
[h(m_1,n), \ h(m_2,n), \ \cdots, \ h(m_{\zeta_{k}},n)] \\ X_k & =
& \left[
\begin{array}{c}
\underline{x}_k(m_1) \\
\underline{x}_k(m_2) \\
\vdots \\
\underline{x}_k(m_{\zeta_k}) \end{array} \right] .\eean  Let \bean
\underline{y}^t_k(n) & = & [y_{(k,1)}(n) \ y_{(k,2)}(n) \ \cdots \
y_{(k,T)}(n)]
\\
\underline{w}^t_k(n) & = & [w_{(k,1)}(n) \ w_{(k,2)}(n) \ \cdots \
w_{(k,T)}(n)] \eean denote the received signal and noise vector at
the $n$th node during the $k$th block.   Then we have \bean
\underline{y}^t_k(n) & = & \underline{h}^t_k(n) X_k \ + \
\underline{w}^t_k(n) . \eean Therefore the totality of the received
signal at the $n$th node up until the end of the $(b-1)$th block is
given by  \bea [\underline{y}^t_1(n) \ \cdots \
\underline{y}^t_{b-1}(n)] & = & [\underline{h}^t_1(n) \
 \cdots \ \underline{h}^t_{b-1}(n)] \notag \\
 & & \left[
\begin{array}{ccc} X_1 & & \\
& \ddots  & \\
& & X_{b-1} \end{array} \right] \notag  \\ & &  \ + \
[\underline{w}^t_1(n) \ \cdots \ \underline{w}^t_{b-1}(n)] .
\label{eq:signal_at_R_n} \eea Note that the vectors
$\underline{h}_l(n)$ as well as the matrices $X_l$, $1 \leq l \leq
b-1$ are in general, of different sizes.

\subsubsection{Signal at Destination}
Since $D \equiv R_{N+1}$, by replacing $n$ by $(N+1)$ and $b-1$ by
$B$ in equation \eqref{eq:signal_at_R_n} above, we recover the
expression for the received signal at the destination during the
$B$th block:
\bea [\underline{y}^t_1(N+1) \ \cdots \ \underline{y}^t_B(N+1)] \notag \\
\ = \ [\underline{h}^t_1(N+1) \
 \cdots \ \underline{h}^t_B(N+1)] \left[
\begin{array}{ccc} X_1 & & \notag \\
& \ddots  & \\
& & X_B \end{array} \right] \\  \ + \ [\underline{w}^t_1(N+1) \
\cdots \ \underline{w}^t_B(N+1)] . \label{eq:signal_at_D} \eea

\subsubsection{Outage of Relay Node} From \eqref{eq:signal_at_R_n}, we note that the
channel ``seen'' by the $n$th relay node over the course of the
first $b-1$ blocks is the MISO (multiple-input single output)
channel characterized by the matrix equation \bea y & = &
[\underline{h}^t_1(n) \
 \cdots \ \underline{h}^t_{b-1}(n)]  \underline{x} \ + \ w .
\label{eq:MISO_channel_at_R_n} \eea

The $n$th relay node can only hope to decode reliably at the end
of the $(b-1)$th block if at that point, it has sufficient mutual
information to recover the transmitted signal whose information
content equals $rB \log (\rho)$ bits. Here $r$ denotes the
multiplexing gain, $\rho$ the signal to noise ratio, and
$r\log(\rho)$ the rate of communication between source and
destination \cite{ZheTse}. If it does not have sufficient
information, then we say that the relay is in outage. Thus the
probability of outage ${\cal P}_{\text{out},n,b-1}(r)$ of the
$n$th relay node at the end of the $(b-1)$th block is given by
\bean {\cal P}_{\text{out},n,b-1}(r) & & \\ \ = \ \text{Pr}\left(
(1 + \rho \sum_{l=1}^{b-1} \mid
 \underline{h}^t_l(n) \mid^2 ) < r\frac{BT}{(b-1)T} \log(\rho) \right) \\
\ = \ \text{Pr}\left( (1 + \rho \sum_{l=1}^{b-1} \mid
 \underline{h}^t_l(n) \mid^2 ) < \frac{rB}{(b-1)} \log(\rho) \right).
\eean Under the DDF protocol, the $n$th relay node at the end of
block $b-1$ uses this expression to decide whether or not it is
ready to decode.  If it is ready to decode, then it will proceed to
do so and then begin transmitting from block $b$ onwards, i.e., $n
\in {\cal I}_b$.

\subsection{Performance
under the DDF Protocol}

In our analysis of the DDF protocol, we will make the assumption
that if a relay does decode erroneously, then this error will
propagate and cause the receiver to decode incorrectly as well. Thus
the receiver at the destination will decode correctly if and only if
in addition to the receiver at the decoder, the receivers at all
intermediate nodes that have participated in relaying of the
transmitted signal have also decoded correctly.

A lower bound on the probability of error of the DDF scheme can thus
be derived by making the assumption that when the channel seen by a
relay node is not in outage and the relay proceeds to decode the
signal transmitted by the source, it will do so without error. Under
this condition, the error probability of the DDF scheme, will be
lower bounded by the probability of outage of the channel
\eqref{eq:signal_at_D}, seen by the destination.  In
Section~\ref{sec:code_construction}, we will construct codes whose
error performance at large SNR is equal to this lower bound, thereby
establishing that this lower bound is indeed the error probability
associated with the DMG tradeoff of the DDF protocol.

Let $\underline{\gamma}$ denote the vector composed of the $ {N+1
\choose 2}$ fading coefficients \[ \left\{h(m,n) \mid \  \\ \ n
> m,  \ \
\begin{array}{c}   1 \leq m \leq N, \\ 2 \leq n \leq (N+1),  \end{array} \right\}
\] ordered lexicographically. We will use $\Gamma$ to denote the
random vector of which $\underline{\gamma}$ is a realization. The
activations sets ${\cal I}_k$ are clearly a function of the channel
realization $\underline{\gamma}$.  Writing ${\cal
I}_k(\underline{\gamma})$ in place of ${\cal I}_k$ to emphasize
this, let us define
\[ {\cal I}(\underline{\gamma}) \ = \
({\cal I}_1(\underline{\gamma}), \cdots, {\cal
I}_B(\underline{\gamma})). \]   Let ${\cal A}$ denote the collection
of all possible activation sets.  It follows that the error
probability of the DDF scheme satisfies \bean {\cal P}_e(r) & \geq &
\sum_{{\cal I} \in {\cal A}} \int_{\underline{\gamma} \in {\cal
R}({\cal I})} p_{\: \Gamma}(\underline{\gamma})
 \ d \underline{\gamma}  \eean where
\[ {\cal R}({\cal I}) \ = \ \left\{ \underline{\gamma} \ \mid \ \begin{array}{c} {\cal
I}(\underline{\gamma}) \ =  {\cal I} \\  \left(1 + \rho \sum_{l=1}^B
\mid
 \underline{h}^t_l(N+1) \mid^2 \right) < r \log(\rho) \end{array}  \right\} .
\]

\subsection{Notation to Aid in Code Analysis}

Returning to the expression for the signal at the $n$th relay node
up until the $(b-1)$th block in \eqref{eq:signal_at_R_n}, we
extend the vectors $\underline{h}_k(n)$ and the matrices $X_k$ to
be of equal size with a view towards the ST code construction to
be presented in Section~\ref{sec:code_construction}.

The vectors \[ \{\underline{h}_k(n) \mid 1 \leq k \leq b-1, \ \ 1
\leq n \leq (N+1) \} \]  will be extended by zero padding, while the
matrices $X_k$, $1 \leq k \leq b-1$will be padded with arbitrary row
vectors. The extra row vectors can be chosen arbitrarily since the
extended matrix $\hat{X}_k$ will be left multiplied by row vectors
$\hat{\underline{h}}_k(n)$ having zeros in the locations
corresponding to the indices of the row vectors where padding of the
matrix $X_k$ takes place.

We thus define, for $1 \leq k \leq b-1$,  \bean \
\hat{\underline{h}}^t_k(n) & = & [\hat{h}_k(1,n) \ \hat{h}_k(2,n)
\ \cdots \hat{h}_k(N,n)] \eean where \bean \hat{h}_k(m,n) & = &
\left\{
\begin{array}{rl} h(m,n) & m \in {\cal
I}_k \\
0 & \text{else} . \end{array} \right. \eean Also, let  \bean
\hat{X}_k & = & \left[
\begin{array}{c}
\hat{\underline{x}}_k(1) \\
\hat{\underline{x}}_k(2) \\
\vdots \\
\hat{\underline{x}}_k(N) \end{array} \right]
 \eean
where \bean \hat{\underline{x}}_k(m) & = & \left\{
\begin{array}{rl} \underline{x}_k(m) & m \in {\cal
I}_k \\
\text{arbitrary $n$-length vector} & \text{else} .
\end{array} \right. \eean

In terms of the extended vector and extended matrix notation, the
received signal at the $n$th relay node, $n \not \in {\cal
I}_{b-1}$ and the destination can respectively be re-expressed in
the form \bea [\underline{y}^t_1(n) \ \cdots \
\underline{y}^t_{b-1}(n)] & = & [\hat{\underline{h}}^t_1(n) \
 \cdots \ \hat{\underline{h}}^t_{b-1}(n)] \notag \\
 & & \left[
\begin{array}{ccc} \hat{X}_1 & & \\
& \ddots  & \\
& & \hat{X}_{b-1} \end{array} \right] \notag  \\ & &  \ + \
[\underline{w}^t_1(n) \ \cdots \ \underline{w}^t_{b-1}(n)] ,
\label{eq:extended_reprsn_signal_at_R_n} \eea
\bea [\underline{y}^t_1(N+1) \ \cdots \ \underline{y}^t_B(N+1)] \notag \\
\ = \ [\hat{\underline{h}}^t_1(N+1) \
 \cdots \ \hat{\underline{h}}^t_B(N+1)] \left[
\begin{array}{ccc} \hat{X}_1 & & \notag \\
& \ddots  & \\
& & \hat{X}_B \end{array} \right] \\  \ + \ [\underline{w}^t_1(N+1)
\ \cdots \ \underline{w}^t_B(N+1)] .
\label{eq:extended_signal_reprsn_signal_at_D} \eea

In this representation, all vectors $\hat{\underline{h}}_l(n)$ are
of the same size, $( 1 \times T)$. The same comment also applies
to the matrices $\hat{X}_l$, $1 \leq l \leq b-1$, which are of
size $(N \times T)$ .

  As will be shown in Section~\ref{sec:optimal_codes_for_DDF} below, ST
codes that are approximately universal for an appropriate class of
block-fading channels will be the building blocks of codes for the
DDF protocol that attain the DMG performance of this channel. For
this reason, a discussion on the block-fading channel is presented
in the next two sections.

\section{The Block-Fading Channel}  \label{sec:block_fading_channel}

\subsection{Outage Probability }

Consider the block-fading MIMO channel with $n_t$ transmit and $n_r$
receive antennas and $B$ blocks, characterized by  \bea
\underline{y}_b & = & H_b \underline{x}_b + \underline{w}_b, \ \ 1
\leq b \leq B. \label{eq:block_fading} \eea Thus each matrix $H_b$
is of size $(n_r \times n_t)$.  The probability of outage of this
channel is given by \bean P_{\text{out}}(r) & \doteq &
\text{Pr}(\sum_{b=1}^B \log
\text{det}(I_{n_r} + \rho H_b H_b^{\dagger}) < rB \log (\rho)) \\
& = & \text{Pr}(\log \text{det}(I_{Bn_r} + \rho \Lambda_H
\Lambda_H^{\dagger}) < rB \log (\rho)) \\
& = & \text{Pr}(\log \text{det}(I_{Bn_t} + \rho \Lambda_H^{\dagger}
\Lambda_H) < rB \log (\rho))  \eean where $\rho$ is the SNR and
where $\Lambda_H$ is the $(Bn_r \times Bn_t)$ block diagonal matrix
\bean \Lambda_H & = &
\left[ \ba{cccc}  H_1 &  & & \\    & H_2 & & \\    &  & \ddots &  \\
 &  & \cdots & H_B   \ea \right] . \eean
 In the above, $\dot =$ and $\dot \leq, \dot \geq$ corresponds to
exponential equality and inequality.  For example,  $y \ \dot = \
\rho^x$ is used to indicate that $\lim\limits_{\rho\rightarrow
\infty} \frac{\log(y)}{\log(\rho)} = x$.  Let $q=n_tB$ and let
\bea
 \lambda_1 \leq \lambda_2 \leq \cdots \leq \lambda_q
 \eea
be an ordering of the $q$ eigenvalues of $\Lambda_H^{\dagger}
\Lambda_H$.  Note that if $n_r
 < n_t$, then
\bea
 \lambda_1 = \lambda_2 = \cdots = \lambda_{(n_t-n_r)B} =0.
 \eea
Let $\delta=([n_t-n_r]B)^{+}$ where $(x)^+$ denotes $\max \{x,0\}$,
and let the $\alpha_i$ be defined by \bean \lambda_i & = &
\rho^{-\alpha_i}, \ \ \delta+1 \leq i \leq q. \eean Then \bean
P_{\text{out}}(r) & = & \text{Pr}(\sum_{i=\delta+1}^{q}
(1-\alpha_i)^{+} < rB ). \eean

We will now proceed to identify a ST code in the next section,
Section~\ref{sec:code_construction}, that is approximately universal
for the class of block-fading channels, i.e., a code that achieves
the D-MG tradeoff of the channel model in (\eqref{eq:block_fading})
for every statistical distribution of the fading coefficients
$\{[H_b]_{i,j}\}$.

Similar construction of codes for such a setting have previously
been identified in \cite{YanBel_Feb_2007,YanBelRek_PerfectParallel}
and independently in \cite{Lu_MB_fading}.  We adopt the
code-construction technique of these papers for the most part,
although the construction presented here is slightly more general,
for example, we permit the individual block codes to be rectangular
and offer flexibility with respect to number of conjugate blocks
employed.  Most importantly though, our proof will establish the new
result that these codes are approximately universal for the
block-fading channel and parallel channels. 

\subsection{Approximately-Universal Codes for the Block-Fading
Channel} \label{sec:code_construction}

\subsubsection{Constructing the Appropriate Cyclic Division
Algebra}

Let $T$ be an integer satisfying $T \geq n_t$.   Let $m \geq B$ be
the smallest integer such that the gcd of $m,T$ equals $1$, i.e.,
$(m,T)=1$. Let $\mathbb{K}, \mathbb{M}$ be cyclic Galois extensions
of $\mathbb{Q}(\imath)$ of degrees $m,T$ whose Galois groups are
generated respectively by the automorphisms $\phi_1$, $\sigma_1$,
i.e., \bean \text{Gal}(\mathbb{K}/\mathbb{Q}(\imath))
& = & <\phi_1> \\
\text{Gal}(\mathbb{M}/\mathbb{Q}(\imath)) & = & <\sigma_1> . \eean
Let $\mathbb{L}$ be the composite of $\mathbb{K}$, $\mathbb{M}$, see
Fig.\ref{fig:CDA_construction}. Then it is known that
$\mathbb{L}/\mathbb{Q}(\imath)$ is cyclic and that further, \bean
\text{Gal}(\mathbb{L}/\mathbb{Q}(\imath)) & \cong &
\text{Gal}(\mathbb{K}/\mathbb{Q}(\imath)) \times
\text{Gal}(\mathbb{M}/\mathbb{Q}(\imath)) .\eean    Thus every
element of $\text{Gal}(\mathbb{L}/\mathbb{Q}(\imath))$ can be
associated with a pair $(\phi_1^i,\sigma_1^j)$ belonging to
$\text{Gal}(\mathbb{K}/\mathbb{Q}(\imath)) \times
\text{Gal}(\mathbb{M}/\mathbb{Q}(\imath))$. Let $\phi,\sigma$ be the
automorphisms associated to the pairs $(\phi_1,\text{id})$,
$(\text{id},\sigma_1)$ respectively.  Then $\phi,\sigma$ are the
generators of the Galois groups $\text{Gal}(\mathbb{L}/\mathbb{M})$,
$\text{Gal}(\mathbb{L}/\mathbb{K})$ respectively.

\begin{figure}[!h]
\begin{center} \hspace*{0.2in}
\xymatrix{ & \mathbb{D} \ar@{-}[d]^T & \\%
& \mathbb{L} \ar@{-}[ld]_{<\sigma>}^T  \ar@{-}[rd]^{<\phi>}_{m} &  \\
\mathbb{K} \ar@{-}[rd]_{<\phi_1>}^{m} & & \mathbb{M} \ar@{-}[ld]^{<\sigma_1>}_{T}  \\
& \Q (\imath)  &
}
\caption{Construction of the underlying
cyclic-division algebra. } \label{fig:CDA_construction}
\end{center}
\end{figure}

Let $\gamma\in \mathbb{K}$ be a non-norm element of the extension
$\mathbb{L}/\mathbb{K}$, i.e., the smallest exponent $e$ for which
$\gamma^e$ is the norm of an element of $\mathbb{L}$ is $T$.  Let
$z$ be an indeterminate satisfying $z^T=\gamma$. Consider the
$T$-dimensional vector space \[D = \{ z^{T-1}\ell_{T-1}\oplus
z^{T-2}\ell_{T-2} \oplus \cdots \ell_0  \ | \ \ell_i\in
\mathbb{L}\}.\] We define multiplication on $D$ by setting $\ell_iz
= z\sigma(\ell_i)$ and extending in a natural fashion. This turns
$D$ into a cyclic division algebra (CDA) whose center is
$\mathbb{K}$ and having $\mathbb{L}$ as a maximal subfield. See
\cite{SetRajSas,EliKumPawKumLu} for an exposition of the relevant
background on division algebras.   Every element $x =
z^{T-1}\ell_{T-1}+z^{T-2}\ell_{T-2}+\cdots +\ell_0$ in $D$ has the
regular representation

\bea X & = & \left[ \begin{array}{cccc}
  \ell_0 & \gamma \sigma (\ell_{T-1}) &  \hdots & \gamma \sigma^{T-1} (\ell_1) \\
  \ell_1 & \sigma (\ell_0)  & \hdots & \gamma \sigma^{T-1} (\ell_2) \\
  \vdots &  \vdots & \ddots & \vdots \\
  \ell_{T-1} & \sigma (\ell_{T-2})  & \hdots  & \sigma^{T-1} (\ell_0) \\
\end{array}
\right] . \label{eq:LeftRegularReprsn} \eea The determinant of
such a matrix is known to lie in $\mathbb{K}$.  Given a matrix $X$
with components $X_{i,j}\in \mathbb{L}$, we define $\phi(X)$ to be
the matrix over $\mathbb{L}$ whose $(i,j)^{th}$ component is given
by $[\phi(X)]_{i,j} = \phi([X]_{i,j}).$  Note that in this case,
\bean
\prod_{i=0}^{m-1} \det(\phi^i(X)) & = & \prod_{i=0}^{m-1} \phi^i(\det(X)) \\
& = & \prod_{i=0}^{m-1} \phi_1^i(\det(X))\\  & \in  &
\mathbb{Q}(\imath).\eean Hence if the elements $\ell_i$ underlying
the matrix $X$ are in addition, restricted to lie in the ring ${\cal
O}_{\mathbb{L}}$ of algebraic integers of $\mathbb{L}$, then we have
that \bean \prod_{i=0}^{m-1} \det(\phi^i(X)) & \in &
\mathbb{Z}(\imath) \eean so that \bea \mid \prod_{i=0}^{m-1}
\det(\phi^i(X)) \mid ^2 & \geq & 1 . \label{eq:nvd_property} \eea

\subsubsection{Space-time Code Construction on the CDA}

Let $\mathcal{X}$ be the rectangular $(n_t \times T)$ ST code
comprised of the first $n_t$ rows of the regular representations of
the elements $\sum_{i=0}^{T-1} z^i \ell_i,$ where $\ell_i$ are
restricted to be of the form: \[ \ell_i =
\sum_{j=1}^T\ell_{i,j}\gamma_j, \ \ \ \ell_{i,j}\in
\mathcal{A}_{\text{QAM}}\] where $\{\gamma_1,\cdots,\gamma_T\}$ are
a basis for $\mathbb{L} / \mathbb{K}$ and where \bean
\mathcal{A}_{\text{QAM}} & = & \left\{a+\imath b \ \mid \ |a|, |b|
\leq (M-1), \ a, b \ \ \text{odd}   \right\} \ \subseteq \
\mathbb{Z}[\imath] \eean denotes the QAM constellation of size
$M^2$. Note that as a result, we have ensured that $\ell_i \in {\cal
O}_{\mathbb{L}}$.  Thus each code matrix in ${\cal X}$ is of the
row-deleted form \bea X & = & \left[
\begin{array}{ccccc}
  \ell_0 & \gamma \sigma (\ell_{T-1}) &  \hdots & \hdots & \gamma \sigma^{T-1} (\ell_1) \\
  \ell_1 & \sigma (\ell_0)  & \hdots & \hdots & \gamma \sigma^{T-1} (\ell_2) \\
  \vdots &  \vdots & \ddots & \ddots & \vdots \\
  \ell_{n_t-1} & \sigma (\ell_{n_t-2})  & \hdots  & \hdots & \gamma \sigma^{T-1} (\ell_{n_t}) \\
\end{array}
\right] . \label{eq:LeftRegularReprsn} \eea

Let ${\cal S}$ be the $(Bn_t\times BT)$ ST code comprised of code
matrices having the
block diagonal form: \[ S = \left\{ \theta \left[  \ba{cccc}  X &  & & \\
 & \phi(X) & & \\    &  & \ddots &  \\    &  &  &
\phi^{B-1}(X) \ea \right], \ \ X\in\mathcal{X} \right\}
\]
where $\theta$ accounts for SNR normalization.  When this code
matrix is in use, the received signal over the block-fading channel
is given by \bea [Y_1 \ Y_2 \ \cdots \ Y_B] & = &  [H_1 \ H_2 \
\cdots \ H_B] S \ + \\
& & \ [W_1 \ W_2 \ \cdots \ W_B].
\label{eq:block-fading_ST_code_channel}\eea

This can also be expressed in the form \bea \left[
\begin{array}{c} Y_1 \\ Y_2 \\ \vdots \\ Y_B \end{array} \right]
& = &  \theta \left[ \ba{cccc}  H_1 &  & & \\    & H_2 & & \\    &  & \ddots &  \\
 &  &  & H_B   \ea \right]
\left[
\begin{array}{c} X \\ \phi(X) \\ \vdots \\ \phi^{B-1}(X) \end{array} \right]
\ + \ \left[
\begin{array}{c} W_1 \\ W_2 \\ \vdots \\ W_B \end{array} \right],
\label{eq:block-fading_ST_code_channel_alternate} \eea in which the
channel matrix is of block-diagonal form.  This latter form is
convenient when comparing the block-fading channel with the parallel
channel.

\subsubsection{Proof of Optimality} \label{subsec:proof of
Optimality}

We will now show that the ST code $\mathcal{S}$ is approximately
universal for the class of channel models described by
(\ref{eq:block_fading}).  From information rate considerations we
must have  $(M^2)^{mT^2} = \rho^{rBT}, $ i.e., $M^2 =
\rho^{\frac{rB}{mT}}$. Also, since  $\theta^2 M^2 \doteq \rho$, we
have $\theta^2\doteq \rho^{1-\frac{rB}{mT}}$.

We next examine the error performance of the code.  Let \bean H & =
& [H_1 \ H_2 \ \cdots \ H_B] \eean and let $\Delta S = S_1-S_2$
where $S_1,S_2,$ are two distinct code matrices belonging to
$\mathcal{S}$.    We have the following expression for the squared
Euclidean distance: \bean d_E^2 &=& \theta^2 Tr\bigl(
 H \Delta S \Delta S^\dag H^\dag \bigr) \\ & = &  \theta^2 Tr\bigl(
 \Lambda_H \Delta S \Delta S^\dag \Lambda_H^\dag \bigr) \no \\ & \geq  & \theta^2 \sum_{i=1}^{n_tB} \lambda_i\mu_i
 \eean by the mismatched eigenvalue bound (see
 \cite{KosWes,EliKumPawKumLu}), where $\mu_1\geq \mu_2\geq\cdots\geq
 \mu_{n_tB}$ are the ordered eigenvalues of $\Delta S \Delta
 S^\dag.$

Let $\Delta \hat{X}$ be the $T \times T$ matrix that corresponds to
$\Delta X$ in the sense that it would have been the matrix obtained
if in constructing the ST code ${\cal X}$, we had not deleted the
bottom $(T-n_t)$ rows from the regular representations of the
elements $\sum_{i=0}^{T-1} z^i \ell_i$.  Correspondingly, let $
\Delta \hat{S}$ be the $(BT \times BT)$ matrix \bean
 \Delta \hat{S} & = &  \left[  \ba{cccc}  \Delta \hat{X} &  & & \\
 & \phi(\Delta \hat{X}) & & \\    &  & \ddots &  \\    &  & \cdots &
\phi^{B-1}(\Delta \hat{X}) \ea \right]. \eean

By the inclusion principle of Hermitian matrices, (see Theorem
4.3.15 of \cite{HorJoh}) the smallest $Bn_t$ eigenvalues of $\Delta
S \Delta S^{\dagger}$ are term-by-term larger than the corresponding
smallest $Bn_t$ eigenvalues of $\Delta \hat{S}  \Delta
\hat{S}^{\dagger}$, i.e.,
\[ \mu_{Bn_t-k+1} \ \geq \ \nu_{BT-k+1} , \ \ \ 1 \leq k \leq Bn_t,
\]
where $\nu_1\geq \nu_2\geq\cdots\geq
 \nu_{BT}$  are the eigenvalues of $\Delta
\hat{S} \Delta \hat{S}^{\dagger}$.

Next, let $\Delta \hat{\hat{S}}$ be the $(mT \times mT)$ matrix
obtained by further extending $\Delta \hat{S}$ to include all $m$
``conjugates'' of $\Delta \hat{X}$, i.e.,  \bean
 \Delta \hat{\hat{S}} & = &  \left[  \ba{cccc}  \Delta \hat{X} &  & & \\
 & \phi(\Delta \hat{X}) & & \\    &  & \ddots &  \\    &  & \cdots &
\phi^{m-1}(\Delta \hat{X}) \ea \right]. \eean Let $\hat{\nu}_i$,
$1 \leq \hat{\nu}_i \leq mT$, be the eigenvalues of $ \Delta
\hat{\hat{S}} [\Delta \hat{\hat{S}}]^{\dagger}$
 ordered such that
\bean \hat{\nu}_i & = & \nu_i, \ \ 1 \leq i \leq BT . \eean

Then for every $1 \leq J \leq q=Bn_t$, we have
\bean d_E^2 &\geq & \theta^2 \sum_{i=0}^{J} \lambda_{q-i} \mu_{q-i} \\
& \geq & \theta^2 \sum_{i=0}^{J} \lambda_{q-i} \nu_{BT-i} \\
& = & \theta^2 \sum_{i=0}^{J} \lambda_{q-i} \hat{\nu}_{BT-i} \\
 & \dot \geq &
\theta^2 \bigl(\prod_{i=0}^{J} \lambda_{q-i} \bigr)^{\frac{1}{J+1}}
\bigl(\prod_{i=0}^{J} \hat{\nu}_{BT-i}
\bigr)^{\frac{1}{J+1}} \\
& \doteq &  \bigl(\prod_{i=0}^{J} \lambda_{q-i}
\bigr)^{\frac{1}{J+1}} \\ & & \left\{ \frac{\prod_{i=1}^{ mT}
\bigl(\theta^2 \hat{\nu}_i \bigr)}{ \prod_{i=1}^{BT-J-1}
\bigl(\theta^2 \hat{\nu}_i \bigr) \prod_{i=BT+1}^{mT}
\bigl(\theta^2 \hat{\nu}_i \bigr) }\right\}^{\frac{1}{J+1}} \\
& \geq & \biggl(\frac{\rho^{-\sum_{i=0}^J \alpha_{q-i}}
(\theta^2)^{mT}}{\rho^{mT-(J+1)}  }\biggr)^{\frac{1}{J+1}} \doteq
\rho^{\frac{\delta_J}{J+1}}
 \eean
where \bean \delta_J &=& mT (1-\frac{rB}{mT})+J+1 - mT -
\sum_{i=0}^J\alpha_{q-i} \\ &=&  J+1-rB -\sum_{i=0}^J \alpha_{q-i} \\
& = & \sum_{i=q-J}^q (1-\alpha_i) - rB . \eean

In this derivation we have made use of the fact that the product
of the eigenvalues is equal to the determinant and of the
non-vanishing determinant property enunciated in
\eqref{eq:nvd_property}. We will now show that if the block-fading
channel is not in outage, that for some $J$, $1\leq J \leq q,$
$\delta_J>0$.

Suppose \beq \sum_{i=1}^{q} (1-\alpha_i)^+\geq (r+\epsilon)B.
\label{eq:alphaFloors1} \eeq

Clearly for some $i$, $\alpha_i <1$.  Since $\alpha_1 \geq \alpha_2
\geq \cdots \geq \alpha_{q}$, assume $\alpha_1 \geq \alpha_2 \geq
\cdots \geq \alpha_{q-J-1} \geq 1, $ and $\alpha_i<1,  \ i \geq
q-J$.  Then (\ref{eq:alphaFloors1}) becomes
\[
\sum_{i=q-J}^q (1-\alpha_i) \  \geq \ (r+\epsilon)B
\]
and this causes the corresponding $\delta_J \geq \epsilon B$ hence
leading to negligible error probability when not in outage.  By
taking the limit as $\epsilon \rightarrow 0$ we see that the
probability of error is negligible in the no-outage region. This
proves that the space-time code ${\cal X}$ achieves the DMG tradeoff
regardless of the statistical distribution of the fading
coefficients that comprise the matrices $\{H_b\}$, i.e., proves
approximate universality of the constructed ST code.

\subsection{Analogous Results Hold for the Parallel Channel}

By parallel channel we will mean the channel given by   \bean \left[
\begin{array}{c} Y_1 \\ Y_2 \\ \vdots \\ Y_B \end{array} \right]
& = & \left[ \ba{cccc}  H_1 &  & & \\    & H_2 & & \\    &  & \ddots &  \\
 &  & \cdots & H_B   \ea \right] S \ + \ \left[
\begin{array}{c} W_1 \\ W_2 \\ \vdots \\ W_B \end{array} \right],
\eean in which the channel matrix is of block-diagonal form.
Consider the $(Bn_t\times T)$
 space-time code ${\cal S}_{\text{par}}$ given by
 \[ {\cal S}_{\text{par}} = \left\{ \theta \left[  \ba{c}  X  \\
  \phi(X)  \\      \vdots   \\
\phi^{B-1}(X) \ea \right], \ \ X\in\mathcal{X} \right\} \] which
when used over the parallel channel leads to the equation below for
the received signal at the receiver,
 \bean \left[
\begin{array}{c} Y_1 \\ Y_2 \\ \vdots \\ Y_B \end{array} \right]
& = & \theta \left[ \ba{cccc}  H_1 &  & & \\    & H_2 & & \\    &  & \ddots &  \\
 &  & \cdots & H_B   \ea \right]
 \left[  \ba{c}  X  \\
  \phi(X)  \\      \vdots   \\
\phi^{B-1}(X) \ea \right]
  \ + \ \left[
\begin{array}{c} W_1 \\ W_2 \\ \vdots \\ W_B \end{array} \right].
\eean  Comparing this equation with the alternate expression for the
block-fading channel given in
\eqref{eq:block-fading_ST_code_channel_alternate} we see that the
expressions are identical.    There is one important difference
though.  In the case of the block-fading channel, a rate requirement
of $R$ bits per channel use translates into a space-time code ${\cal
S}$ of size $2^{RBT} \ = \ \rho^{rBT}$, whereas in the case of the
parallel channel, the size of the corresponding ST code ${\cal
S}_{\text{par}}$ is required to be $2^{RT}\ = \ \rho^{rT}$.

It follows from this that by replacing $rB$ by $r$, one can
similarly prove approximate universality of the code ${\cal
S}_{\text{par}}$ for the class of parallel channels.  We omit the
details.

\subsection{DMT-optimal Codes for the General MIMO-OFDM Channel}

The MIMO-OFDM channel can be regarded as a parallel channel in which
each parallel block corresponds to a different subcarrier and can
thus be represented in the form: \bean \underline{y}_i & = & \theta
H_l \underline{x}_l + \underline{w}_l , \ \ 1 \leq l \leq Q, \eean
where where $Q$ is the number of OFDM tones or sub-carriers
\cite{CorBol}.  The matrices $H_l$ are correlated in general, with a
correlation derived from the time-dispersion of the original ISI
channel.  Since the code ${\cal S}_{\text{par}}$ is approximately
universal, this means that the code  ${\cal S}_{\text{par}}$ is DMG
optimal when used over the MIMO fading channel.    When the code
${\cal S}_{\text{par}}$ is used over the MIMO-OFDM channel, the
received-signal equation will take on the form
 \bean \left[
\begin{array}{c} Y_1 \\ Y_2 \\ \vdots \\ Y_Q \end{array} \right]
& = & \theta \left[ \ba{cccc}  H_1 &  & & \\    & H_2 & & \\    &  & \ddots &  \\
 &  & \cdots & H_Q   \ea \right]
 \left[  \ba{c}  X  \\
  \phi(X)  \\      \vdots   \\
\phi^{Q-1}(X) \ea \right]
  \ + \ \left[
\begin{array}{c} W_1 \\ W_2 \\ \vdots \\ W_Q \end{array} \right].
\eean

DMT-optimal codes for the OFDM channel have previously been
constructed in \cite{YanBelRek_PerfectParallel} and
\cite{Lu_ITW_Bergen}.  In \cite{YanBelRek_PerfectParallel}, the
authors provide a proof only for the case when the matrices $H_n$
appearing along the diagonal are i.i.d. Rayleigh.  The DMT-optimal
construction in \cite{Lu_ITW_Bergen} is for the SIMO-OFDM case. We
thus believe the results in this paper represent the first
construction of DMT-optimal codes for the general OFDM-MIMO channel.

\section{Codes Attaining the DMG of the DDF Protocol}
\label{sec:optimal_codes_for_DDF}

We now show how ST codes constructed for the block-fading channel
can be used to construct optimal codes under the DDF protocol.

We consider the DDF protocol as it applies to a communication
system in which there are a total of $N+1$ nodes that cooperate in
the communication between source node $S$ and destination node
$D$.

As in Sections~\ref{sec:introduction},\ref{sec:DDF_protocol}, under
the DDF protocol, the source transmits for a total time duration of
$BT$ channel uses.   This collection of $BT$ channel uses is
partitioned into $B$ blocks with each block composed of $T$ channel
uses.  Communication is slotted in the sense that each relay is
constrained to commence transmission only at block boundaries. A
relay will begin transmitting after listening for a time duration
equal to $b$ blocks only if the channel ``seen'' by the relay is
good enough to enable it to decode the  signal from the source with
negligible error probability.

Our coding strategy runs as follows.  The role played by $n_t$ in
the block-fading scenario is now played by the number $N$ which is
the number of nodes in the network capable of transmitting to the
destination. Let $\mathcal{X}$ be the rectangular $(N \times T)$
ST code comprised of the first $N$ rows of the regular
representations of the elements $\sum_{i=0}^{T-1} z^i \ell_i,$
where $\ell_i$ are restricted to be of the form: \[ \ell_i =
\sum_{j=1}^T\ell_{i,j}\gamma_j, \ \ \ \ell_{i,j}\in
\mathcal{A}_{\text{QAM}} . \]    Let $\mathcal{D}$ be the
$(BN\times BT)$ ST code comprised of code matrices having the
block diagonal form: \beq {\cal D} = \left\{ \theta \left[  \ba{cccc}  X &  & & \\
 & \phi(X) & & \\    &  & \ddots &  \\    &  & \cdots &
\phi^{B-1}(X) \ea \right], \ \ X\in\mathcal{X} \right\}
\label{eq:code_matrix_form}\eeq where $\theta$ accounts for SNR
normalization. The code to be used then has the following simple
description.  The source $S$ sends the first row of each of the
matrices $X$, $\phi(X)$, $\cdots$, $\phi^{B-1}(X)$ in successive
blocks. Let us assume that relay node $R_n$, $2 \leq n \leq N$, is
not in outage for the first time at the conclusion of the
$(b-1)$th block.  Then $R_n$ is ready to decode a the end of the
$b-1$th block. Thereafter, it proceeds to send in succession, the
$n$th rows of the matrices $\phi^{b}(X)$, $\phi^{b+1}(X)$,
$\cdots$, $\phi^{B-1}(X)$.  Thus the appended matrices $\hat{X}_i$
appearing in \eqref{eq:extended_reprsn_signal_at_R_n},
\eqref{eq:extended_signal_reprsn_signal_at_D}, correspond to the
matrices $\phi^{i-1}(X)$ in \eqref{eq:code_matrix_form}.

It is easy to show using the results stated earlier relating to the
block-fading channel that this coding strategy ensures that whenever
a relay node decodes, it does so with negligible probability of
error.  The destination error probability is also similarly
guaranteed to have error probability that is SNR-equivalent to the
outage probability, thus proving DMG-optimality of the constructed
ST code.

This follows since each relay node $R_n, \ n \not \in {\cal
I}_{b-1}$ ``sees'' a block-fading channel (see
\eqref{eq:extended_reprsn_signal_at_R_n} and the coding strategy
we have adopted ensures that the code matrix carrying data from
the nodes in ${\cal I}_{b-1}$ is DMT optimal for the corresponding
block-fading channel.  A similar statement is true for the relay
$R_{N+1}$ that corresponds to the destination, since the
corresponding channel equation is of the same block-fading form
see \eqref{eq:extended_signal_reprsn_signal_at_D}.

\subsection{Example}

We illustrate with an example.  Consider a network in which there
are a total of $N+1=5$ nodes including source $S \equiv R_1$ and
destination $D \equiv R_{5}$ nodes.  Let the block length $T=4$ and
the number of blocks $B=4$.  Let us assume that at the end of the
first block, relay $R_3$ is not in outage and therefore in a
position to decode.  Let us assume that relay $R_4$ is ready to
decode at the end of the $3$rd block and that relay $R_2$ is in
outage throughout and thus does not participate in the
communication. Thus  \bean {\cal I}_1 & = &
\{1\}, \\
{\cal I}_2 & = & \{1,3\} ,\\
{\cal I}_3 & = & \{1,3\}, \\
{\cal I}_4 & = & \{1,3,4\}
 \eean
here.    In terms of the notation introduced in
Section~\ref{subsec:notation}, the signal received by relays
$R_2,R_3,R_4,R_5$ prior to decoding are given as follows.
Decisions at each of the relays $R_3,R_4,R_5$ are based on the
received signals \bean \underline{y}_1^t(3) & = &
h(1,3)\underline{x}_1^t(1) \ + \ \underline{w}_1^t(3) , \eean

\bean [\underline{y}_1^t(4) \ \underline{y}_2^t(4) \
\underline{y}_3^t(4)  ] & = &   [h(1,4) \ h(1,4) \ h(3,4) \ h(1,4) \
h(3,4) ] \\ & & \left[
\begin{array}{ccc} \underline{x}_1^t(1) & &
\\
& \underline{x}_2^t(1) &  \\
& \underline{x}_2^t(3) &  \\
& & \underline{x}_3^t(1)   \\
& & \underline{x}_3^t(3)  \\
\end{array} \right]  \\ & &  +
[\underline{w}_1^t(4) \ \underline{w}_2^t(4)\ \underline{w}_3^t(4)]
 \eean

\bean  [ \underline{y}_1^t(5) \ \underline{y}_2^t(5) \
\underline{y}_3^t(5) \ \ \underline{y}_4^t(5) ] \ = \  & & \\
\left[ h(1,5) \ h(1,5) \ h(3,5) \ h(1,5) \ h(3,5) \ h(1,5) \ h(3,5)
\ h(4,5) \right] & & \\
 \left[
\begin{array}{cccc} \underline{x}_1^t(1) & & & \\
& \underline{x}_2^t(1) & & \\
& \underline{x}_2^t(3) & & \\
& & \underline{x}_3^t(1) &  \\
& & \underline{x}_3^t(3) &  \\
& & & \underline{x}_4^t(1)  \\
& & & \underline{x}_4^t(3)   \\
& & & \underline{x}_4^t(4)
\end{array} \right]   & & \\   +
\left[ \underline{w}_1^t(5) \ \underline{w}_2^t(5)\
\underline{w}_3^t(5) \ \ \underline{w}_4^t(5) \right] & &
 \eean
respectively.

Set \bean X(1) & := & \left[ \begin{array}{c} \underline{x}_1(1)
\\ \underline{x}_1(2) \\ \underline{x}_1(3) \\ \underline{x}_1(4)
\end{array} \right] \ = \
 \left[ \begin{array}{cccc}
  \ell_0 & \gamma \sigma (\ell_{3}) &  \gamma \sigma^2 (\ell_{2}) & \gamma \sigma^{3} (\ell_1) \\
  \ell_1 & \sigma (\ell_0)  & \gamma \sigma^2(\ell_3) & \gamma \sigma^{3} (\ell_2) \\
 \ell_2 & \sigma (\ell_1)  & \sigma^2(\ell_0) & \gamma \sigma^{3} (\ell_3) \\
  \ell_{3} & \sigma (\ell_{2})  & \sigma^2 (\ell_{1}) & \sigma^{3} (\ell_0) \\
\end{array}
\right]  \\
X(2) & := &  \phi \left\{ X(1) \right\} \\
X(3) & := &  \phi^2 \left\{ X(1) \right\} \\
X(4) & := &  \phi^3 \left\{ X(1) \right\} .
 \eean
The corresponding extended vectors are given by   \bean
[\hat{\underline{h}}^t_1(3)] & = & [h(1,3) \ 0 \ 0 \ 0 ] \eean

\bean
\hat{\underline{h}}^t_1(4) & = & [h(1,4) \ 0 \ 0 \ 0] \\
\hat{\underline{h}}^t_2(4) & = & [h(1,4) \ 0 \ h(3,4) \ 0] \\
\hat{\underline{h}}^t_3(4) & = & [h(1,4) \ 0 \ h(3,4) \ 0] \eean

\bean
\hat{\underline{h}}^t_1(5) & = & [h(1,5) \ 0 \ 0 \ 0] \\
\hat{\underline{h}}^t_2(5) & = & [h(1,5) \ 0 \ h(3,5) \ 0] \\
\hat{\underline{h}}^t_3(5) & = & [h(1,5) \ 0 \ h(3,5) \ 0] \\
\hat{\underline{h}}^t_1(5) & = & [h(1,3) \ 0 \ h(3,5) \ h(4,5) ]
\eean and we have  \bean [\underline{y}_1^t(5) \
\underline{y}_2^t(5)\ \underline{y}_3^t(5) \ \ \underline{y}_4^t(5)
] & = & [ \hat{\underline{h}}^t_1(5) \ \hat{\underline{h}}^t_2(5)  \
\hat{\underline{h}}^t_3(5)  \ \hat{\underline{h}}^t_4(5)] \left[
\begin{array}{cccc} X(1) & & & \\ & X(2) & & \\ & & X(3) & \\ & &
& X(4) \end{array} \right] . \eean

\subsection{Extension to Multiple Antenna Case}

The extension to the case of multiple antennas at each relay node
and at the destination is straightforward.  

\section{An Alamouti-based Code for the case of a Single Relay}

In this section, we provide a particularly simple code
construction for the case when in addition to the source $S$ and
destination $D$, there is a single relay antenna $R_2$.  The basic
building block for the distributed space-time code is an Alamouti
code.  A separate proof is given here as the proof given in
previous sections does not apply here, primarily because the
Alamouti code is not an approximately universal code\footnote{This
result was first presented at a poster session in Allerton 2006}.

\subsection{Constructing the Appropriate Cyclic Division Algebra}

The cyclic division algebra is constructed along the same lines as
before, with some differences, for example $\mathbb{Q}$ here plays
the role of $\mathbb{Q}(\imath)$ earlier .

Here the number of channel uses in each block equals $2$, i.e.,
$T=2$.  Let $m \geq B$ be the smallest integer such that
$(m,T)=1$, i.e., $m$ is the smallest odd integer $\geq B$.  Set
$\mathbb{M}=\mathbb{Q}(\imath)$ and let $\mathbb{K}$ be a cyclic
Galois extension of $\mathbb{Q}$ of degree $m$. Let the Galois
groups of $\mathbb{K}/\mathbb{Q}$ and $\mathbb{M}/\mathbb{Q}$ be
generated respectively by the automorphisms $\phi_1$, $\sigma_1$,
i.e., \bean \text{Gal}(\mathbb{K}/\mathbb{Q})
& = & <\phi_1> \\
\text{Gal}(\mathbb{M}/\mathbb{Q}) & = & <\sigma_1> . \eean Thus
$\sigma_1$ corresponds to the complex conjugation operator. Let
$\mathbb{L}$ be the composite of $\mathbb{K}$, $\mathbb{M}$, see
Fig.\ref{fig:CDA_construction_single_relay}.

\begin{figure}[!h]
\begin{center} \hspace*{0.2in}
\xymatrix{ & \mathbb{D} \ar@{-}[d]^2 & \\%
& \mathbb{L} \ar@{-}[ld]_{<\sigma>}^2  \ar@{-}[rd]^{<\phi>}_{m} &  \\
\mathbb{K} \ar@{-}[rd]_{<\phi_1>}^{m} & & \mathbb{M}=\mathbb{Q}(\imath) \ar@{-}[ld]^{<\sigma_1>}_{2}  \\
& \Q   & } \caption{Construction of the underlying cyclic-division
algebra. } \label{fig:CDA_construction_single_relay}
\end{center}
\end{figure}

Then it is known that \bean
\text{Gal}(\mathbb{L}/\mathbb{Q}(\imath)) & \cong &
\text{Gal}(\mathbb{K}/\mathbb{Q}(\imath)) \times
\text{Gal}(\mathbb{M}/\mathbb{Q}(\imath)) .\eean    Thus every
element of $\text{Gal}(\mathbb{L}/\mathbb{Q}(\imath))$ can be
associated with a pair $(\phi_1^i,\sigma_1^j)$ belonging to
$\text{Gal}(\mathbb{K}/\mathbb{Q}(\imath)) \times
\text{Gal}(\mathbb{M}/\mathbb{Q}(\imath))$. Let $\phi,\sigma$ be
the automorphisms associated to the pairs $(\phi_1,\text{id})$,
$(\text{id},\sigma_1)$ respectively.  Then $\phi,\sigma$ are the
generators of the Galois groups
$\text{Gal}(\mathbb{L}/\mathbb{M})$,
$\text{Gal}(\mathbb{L}/\mathbb{K})$ respectively. Here again,
$\sigma$ is the complex conjugation operator operating on
$\mathbb{L}$. We note that $\sigma$ commutes with $\phi$.

Let $\gamma=-1$.  Then $\gamma$ is a non-norm element of the
extension $\mathbb{L}/\mathbb{K}$, i.e., the smallest exponent $e$
for which $\gamma^e$ is the norm of an element of $\mathbb{L}$ is
$2$. This follows because the norm in $\mathbb{L}/\mathbb{K}$ of
any element in $\mathbb{L}$ is non-negative.  Let $z$ be an
indeterminate satisfying $z^2=\gamma$. Consider the
$2$-dimensional vector space
\[D = \{ z\ell_{1}\oplus \ell_0  \ | \ \ell_i\in \mathbb{L}\}.\]
We define multiplication on $D$ by setting $\ell_iz =
z\sigma(\ell_i)$ and extending in a natural fashion. This turns
$D$ into a CDA whose center is $\mathbb{K}$ and having
$\mathbb{L}$ as a maximal subfield.  Every element $a =
z\ell_1+\ell_0$ has the regular representation \bea X & = & \left[
\begin{array}{cc}
  \ell_0 & -\ell_0^{*} \\
  \ell_1 & \ell_0^{*} \end{array} \right]  \label{eq:LeftRegularReprsn} \eea
  which we recognize as the familiar Alamouti
  code matrix. The determinant of
such a matrix is clearly real and thus lies in $\mathbb{K}$. Note
also that the rows of every such matrix are orthogonal and hence
the two eigenvalues of the matrix have the same magnitude.

Given a matrix $X$ with components $X_{i,j}\in \mathbb{L}$, we
define $\phi(X)$ to be the matrix over $\mathbb{L}$ whose
$(i,j)^{th}$ component is given by $[\phi(X)]_{i,j} =
\phi([X]_{i,j}).$  Note that in this case, \bean
\prod_{i=0}^{m-1} \det(\phi^i(X)) & = & \prod_{i=0}^{m-1} \phi^i(\det(X)) \\
& \in  & \mathbb{Q}.\eean Hence if the elements $\ell_i$
underlying the matrix $X$ are in addition, restricted to lie in
the ring ${\cal O}_{\mathbb{L}}$ of algebraic integers of
$\mathbb{L}$, then we have that \bean \prod_{i=0}^{m-1}
\det(\phi^i(X)) & \in &   \mathbb{Z} \eean so that \bea \mid
\prod_{i=0}^{m-1} \det(\phi^i(X)) \mid ^2 & \geq & 1 .
\label{eq:nvd_property_single_relay} \eea

\subsection{Channel Models and Outage}
\label{sec:ch_model_outage_single_relay}

We assume as before that the total duration of communication is
$2B$ channel uses, partitioned into $B$ blocks of $2$ channel uses
each. Let us assume that at the end of the $(b-1)$th block, the
relay determines for the first time that it is not in outage and
begins to transmit from block $b$ onwards. The channel perceived
by the relay antenna is given by \bean y & = & h_{sr} x + w  \eean
where $w$ is the usual additive noise.   The probability of outage
of this channel is given by \bean \text{Pr}(\log(1 + \rho \mid
h_{sr} \mid^2)) < \frac{2rB}{2(b-1)} \log (\rho)  \\
= \text{Pr}( (1-\beta_1)^{+} < \frac{rB}{(b-1)}) \eean where
$\beta_1$ is defined by \bean \mid h_{sr} \mid^2 & \doteq &
\rho^{-\beta_1} . \eean

The channel seen by the destination takes on the form \bean y & =
& [\underbrace{h_{sr} \ \cdots \ h_{sr}}_{\text{$(b-1)$ terms}} \
\underbrace{h_{sr} \ h_{rd} \ \cdots \ h_{sr} \
h_{rd}}_{\text{$2[B-(b-1)]$ terms}} ]\underline{x} + w \\
& = & \underline{h}^t \underline{x} + w , \eean where we have
defined \bea \underline{h} & := & [h_{sr} \ \cdots \ h_{sr}\
h_{sr} \ h_{rd} \ \cdots \ h_{sr} \ h_{rd} ]. \label{eq:h_defn}
\eea

\subsection{DMT-Optimal Code Construction}

Optimal code construction proceeds as follows.  The source
transmits the $B$ blocks \bean \theta [A \ \phi(A) \ \cdots
\phi^{B-1}(A)] \eean in succession, where \bean A & = & [\ell_0 \
-\ell_1^{*} ] \eean and where \bean \ell_i & = & \sum_{j=1}^{m}
\ell_{ij} \gamma_j \eean where $\gamma_j$ is a basis for
$\mathbb{L}/\mathbb{Q}(\imath)$ and where $\ell_{ij} \in {\cal
A}_{\text{QAM}} $.

The relay transmits from block $b$ onwards and its transmissions
are of the form \bean \theta [\phi^b(C) \ \phi^{b+1}(C) \ \cdots \
\phi^{B-1}(C) ] \eean where \bean C & = & [\ell_1 \ \ \ \ell_0^{*}
] . \eean The signal seen by the receiver at the destination is
thus of the form \bean \underline{y} & = & \theta \underline{h}
\left[
\begin{array}{ccccccc} A & & & & & & \\
& & \ddots & & & & \\
& & & \phi^{b-1}(A) & & & \\
& & & & \phi^b(A) & & \\
& & & & \phi^b(C) & & \\
& & & & & \ddots &  \\
& & & & & &  \phi^{B-1}(A) \\
& & & & & &  \phi^{B-1}(C) \end{array} \right] +\underline{w} ,
\eean with $\underline{h}$ given in \eqref{eq:h_defn}.

\subsection{Proof of Optimality}

We will show as in Section \ref{sec:optimal_codes_for_DDF} that
when either the relay or the destination is not in outage, the
error probability incurred by this code is negligible i.e., of
order $\rho^{-\infty}$ thus proving DMT optimality of the code.

Note that from rate considerations, we must have \bean (M^2)^{2m}
& = &
\rho^{2 r B} \\
\therefore M^2 & = & \rho^{\frac{rB}{m}} \\
\theta^2 M^2 & \doteq & \rho \\
\Rightarrow \theta^2 & \doteq & \rho^{1-\frac{rB}{m}} . \eean

\subsubsection{Optimality in the Broadcast Phase}

Let \bean {\cal A}_{\text{$\Delta$QAM}} & = & \left\{a+ \imath b
\mid a,b, \text{even},  \ \ 0 \leq |a|, |b| \leq 2(M-1) \right\}.
\eean

The Euclidean distance between the received matrices associated to
code matrices $X_1$, $X_2$ is given by \bean d_{E}^2(X_1, X_2) & =
& \mid h_{sr} \mid^2 \theta^2 \sum_{i=0}^{b-1}\left[ \mid
\phi^i(\ell_0)\mid^2 + \mid \phi^i(-\ell_1^{*}) \mid^2 \right], \
\ \ell_i \in {\cal A}_{\Delta \text{QAM}} \\ & = & \mid h_{sr}
\mid^2 \theta^2 \sum_{i=0}^{b-1} \phi^i( \mid \ell_0 \mid^2 + \mid
\ell_1
\mid^2) \\
& = & \mid h_{sr} \mid^2 \theta^2 \sum_{i=0}^{b-1}\phi^i(\ell) , \
\ \ \ \ell = \mid \ell_0 \mid^2 + \mid \ell_1
\mid^2 \\
& \geq & \mid h_{sr} \mid^2 \theta^2 b \left[ \prod_{i=0}^{b-1}
 \phi^i(\ell) \right]^{\frac{1}{b-1}} \\
& \doteq  & \mid h_{sr} \mid^2 \left[ \frac{\prod_{i=0}^{m-1}
\theta^2 \phi^i(\ell)}{\prod_{i=b}^{m-1}\theta^2 \phi^i (\ell) }
\right]^{\frac{1}{b-1}} \\
& = & \mid h_{sr} \mid^2 \left(
\frac{\rho^{m(1-\frac{rB}{m})}}{\rho^{(m-b+1)}}
\right)^{\frac{1}{b-1}} \\
& = & 
\rho^{1-\beta_1-\frac{rB}{b-1}} . \eean

Consider the no-outage region associated to rate $r+\epsilon$:
\bean (1-\beta_1)^{+} & \geq & (r + \epsilon) \frac{rB}{b-1} \eean
and it follows that the probability of error is negligible for all
$\epsilon >0$.

\subsubsection{Code in the Cooperation Phase}

Here, by making use of the orthogonality of the rows of the
Alamouti code, we can bound the minimum Euclidean distance as
follows: \bean d^2_{E}(\Delta X) & \geq & \theta^2 |h_{sd}|^2
\sum_{i=0}^{b-1} \phi^i(|\ell_0|^2+|-\ell_1|^2) \\
& & +\theta^2 (|h_{sd}|^2+|h_{rd}|^2) \sum_{i=b}^{B-1}
\phi^i(|\ell_0|^2+|-\ell_1|^2) \\
& = &  \theta^2 |h_{sd}|^2 \sum_{i=0}^{B-1} \phi^i(\ell) +
\theta^2 (|h_{rd}|^2) \sum_{i=b}^{B-1} \phi^i(\ell) \\
& = & \theta^2  \mid h_{sd} \mid^2  \sum_{i=0}^{b-1} \phi^i(\ell)
\ + \ \theta^2 [\mid h_{sd} \mid^2 + \mid h_{rd} \mid^2]
\sum_{i=b}^{B-1} \phi^i(\ell). \eean Let us abbreviate and write
\bean h_1 & = & h_{sr} \\
h_2 & = & \sqrt{\mid h_{sr} \mid^2 + \mid h_{rd} \mid^2} \\
\underline{h}^{'} & = & [\underbrace{h_1 \cdots h_1}_{b-1
\text{terms}} \underbrace{h_2  \cdots h_2}_{B-b+1 \text{terms}}]
\eean

Then this squared Euclidean distance is also the squared Euclidean
distance over the block-fading channel shown below: \bean
\underline{y} & = & \theta \underline{h}^{'} \left[
\begin{array}{cccc} \ell & & &  \\
& \phi(\ell) & &  \\
& & \ddots &  \\
& & & \phi^{B-1}(\ell)  \end{array} \right] +\underline{w} . \eean
Note that since \bean \mid \underline{h} \mid^2 & = & \mid
\underline{h}^{'} \mid^2 \eean both channels are in outage for
precisely the same set of values of the fading coefficients
$h_{sd}$, $h_{rd}$.  On the other hand by our results in
Section~\ref{subsec:proof of Optimality}, the block diagonal code
appearing in the equation above has negligible error probability
when the channel is not in outage. This proves DMT optimality of the
code in the cooperation phase as well.

\begin{center}
{\bf Acknowledgement}
\end{center}

The authors would like to thank Hsiao-feng (Francis) Lu and
Jean-Claude Belfiore for useful discussions.

\bibliographystyle{IEEEbib}

\end{document}